\documentclass{article}

\usepackage{arxiv}

\usepackage[utf8]{inputenc} 
\usepackage[T1]{fontenc}    
\usepackage{hyperref}       
\usepackage{url}            
\usepackage{booktabs}       
\usepackage{amsfonts}       
\usepackage{nicefrac}       
\usepackage{microtype}      
\usepackage{lipsum}
\usepackage{graphicx}
\graphicspath{ {./images/} }
\usepackage{amsmath}
\usepackage{float}

\title{Comparison of Mathematical Models for Subscription Services Using Optimization Problems and Quantum Information Theory

\small Feasibility of Implementing Optimization Problem Algorithms on Quantum Computers}

\author{
 Misao Fukuda \\
  \\
  \\
  \texttt{fdk.sm01@gmail.com} 
  \\
}

\begin{document}
\maketitle
\begin{abstract}
The purpose of this research is to explore whether it is possible to construct a design theory for subscription services for intangible goods from a time discounting perspective, based on quantum information theory, which is the foundational theory for quantum computers and similar technologies. To this end, we propose a mathematical model of subscription services using optimization problems based on optimal growth theory from standard economics, and with reference to microeconomics, we define utility as a value function of customer satisfaction derived from quantum mutual information, an entropy measure in quantum information theory, by considering time discounting. We propose the quantification of customer satisfaction and the formulation of consumer surplus. In the mathematical model of subscription services, the existence of a minimum value in the time-discounted customer satisfaction value function under budget constraints, and the realization of a mathematical expression for consumer surplus, could be explained by the laws of behavioral economics. This yielded new insights into the design of individually customized customer experiences, enhanced the feasibility of constructing economic models based on quantum information theory and the mathematical design of customer experiences, raised the possibility that mathematical models using quantum information theory can achieve greater economic welfare than standard economics, and increased the feasibility of implementing optimization problem algorithms on quantum computers.
\end{abstract}



\section{Introduction}
The significance of this research lies in suggesting solutions to the propositions of the feasibility of constructing a design theory for subscription services through mathematical models from a time-discounting perspective, and the feasibility of mathematically designing customized customer experiences through these mathematical models, by examining intangible goods based on quantum information theory.

Examples of research on customers and consumers in subscription services include those related to sales and customer lifetime value (McCarthy et al. 2017), service continuation and churn prevention (retention) (Ascarza et al. 2018), prevention of defection from contracts and reactivation of dormant customers (Kanuri and Andrews 2019), and other studies focusing primarily on economic and monetary value or contracts themselves. However, there are almost no examples of discussions aimed at characterizing customer satisfaction in subscription services, and the extension of existing theories and organization of characteristics can be considered an important research topic. The lack of research on customer satisfaction with the continuous consumption of digital goods provided by subscription services is also identified as the first item in Future Research by Micken et al. (2020), indicating that this is a globally recognized research issue (Dazai 2022).

Quantum information theory, which is the foundational theory for quantum computers and similar technologies, is based on the measurement problem, which states that the state of an object changes through observation (Nielsen and Chuang 2000). Quantum decision theory applies this quantum information theory to social sciences, and research has been conducted on the mathematical structure of behavioral economics that represents the irrational decision-making of humans by Tversky and Kahneman (1992) (Cheon and Takahashi 2010, Takahashi 2013, Yukalov and Sornette 2017, Fukuda 2022, 2023, 2024). Fukuda (2022) set up and analyzed an economic model of intangible goods based on the observation theory of quantum information. By deriving the value function of prospect theory and finding correspondences with the laws of behavioral economics that violate the independence axiom, the quantum information approach increased the possibility there being a candidate for both an economic model for design theory of intangible goods and a mathematical design model of customer experience. From Fukuda (2023, 2024), by adding knowledge about the mathematical design of nudges as individually customized customer experiences, it became possible to make the mathematical structure of total social surplus larger than that in standard economics. This increased the feasibility of economic models based on quantum information theory and the mathematical design of customized nudges. However, the results of deriving the quantification of customer satisfaction and social welfare functions for subscription services based on quantum information theory in this economic model, and of deriving the optimal timing for interventions through nudges to increase repeat rates, have not yet been obtained.

For comparison with mathematical models using quantum information theory, we set up a mathematical model of subscription services as an optimization problem using optimal growth theory based on standard economics with exponential discounting, which organizes the extension of existing theories and their characteristics, define a value function of customer satisfaction, and seek the minimum condition of this customer satisfaction value function under budget constraints. We furthermore set up a mathematical model using quantum information theory. Microeconomics derives consumer surplus from the utility maximization condition under budget constraints and the expenditure minimization condition for a given level of utility under budget constraints. Referring to this, we define the budget constraint equation with demand quantity as contract time, and by considering time discounting, we define utility as a value function of customer satisfaction derived from quantum mutual information—an entropy measure in quantum information theory—and seek the minimum condition of this customer satisfaction value function under budget constraints. Then, in both the standard economics and quantum information theory mathematical models that we have set up, we verify the correspondence between mathematical properties such as consumer surplus derived from the expenditure minimization condition for a given customer satisfaction value function and the laws of behavioral economics.

The aim of this research is to obtain new insights into consumer surplus and the design of individually customized customer experiences in subscription services by quantifying customer satisfaction using quantum information theory based on conventional microeconomics, comparing with insights obtained from the analysis results of a mathematical model of subscription services using optimization problems set up based on standard economics, in response to unresolved issues in prior research.

As the proportion of intangible goods in industry has increased rapidly in recent years, there is an urgent need to develop scientific methods for handling intangible goods, rather than relying on experience and intuition. The necessity of attempts such as this research is also increasing.

As a result of this research, it became clear that in both the standard economics and quantum information theory mathematical models, a minimum value exists in the time-discounted customer satisfaction value function under budget constraints, and that in the quantum information theory mathematical model, customized customer experiences can be designed. This revealed that in subscription services, the mathematical model using quantum information theory can express more than the mathematical model using optimization problems in standard economics.

The new insights into the quantification of customer satisfaction using quantum information theory and the consumer surplus derived from it provided by this research enhance the feasibility of the mathematical design of customer experience and may contribute to the further development of today’s economy in which the proportion of intangible goods is increasing.

\begin{figure}[H]
    \centering
    \includegraphics[width=1\linewidth]{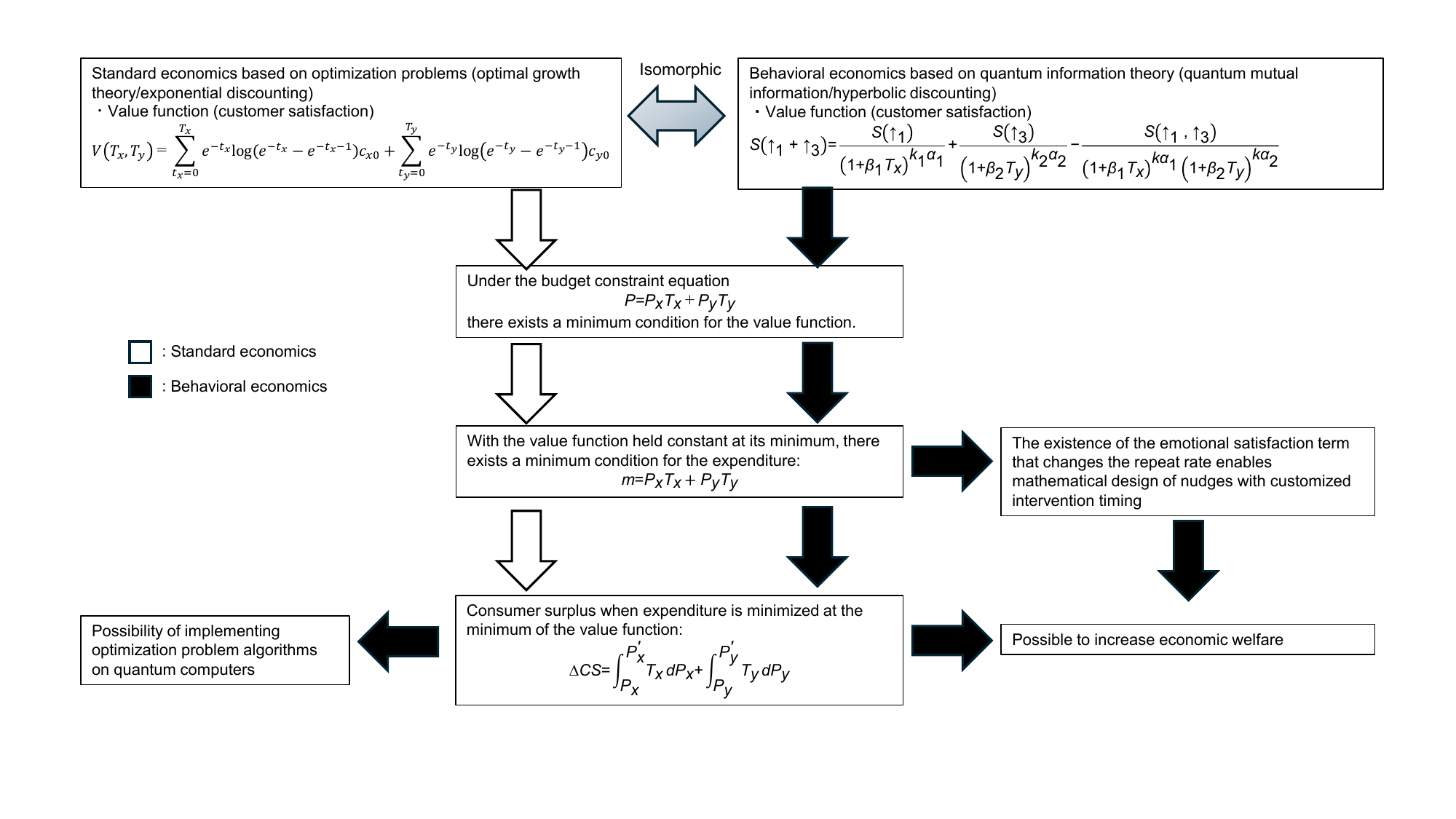}
    \caption{Overview of the mathematical system for subscription services based on optimization problems and quantum information theory, with reference to a microeconomics framework}
    \label{fig:placeholder}
\end{figure}

\section{MATHEMATICAL MODELS OF SUBSCRIPTION SERVICES}
Table 1 shows the issues examined in this research and the solution methods discussed below.

\label{sec:headings}

\begin{table}[ht]
\centering
\caption{Research Issues and Solution Methods}
 \begin{tabular}{|p{7.0cm}|p{7.0cm}|}
   \hline
   Issue & Solution Method \\
   \hline
   A mathematical representation of subscription services using standard economics has not been implemented & Based on optimal growth theory, customer satisfaction is derived from two utility functions assumed to be logarithmic functions using dynamic programming, which is an optimization problem, and minimized under budget constraint conditions \\
   \hline
   A mathematical representation of subscription services using quantum information theory has not been implemented & Based on quantum information theory, two mutually correlated customer satisfactions are expressed with time discounting and minimized under budget constraint conditions \\
   \hline
   A mathematical representation of customized customer experience has not been implemented & The coefficients of the basis are obtained from the minimum condition as a function of time for the perception of the supply and demand process of intangible goods when the value function becomes minimum at a certain contract time, and the contract time at which this minimum value occurs is minimized \\
   \hline
   \end{tabular}   
\end{table}

\subsection{MATHEMATICAL MODEL USING STANDARD ECONOMICS}

\subsubsection{OPTIMAL GROWTH THEORY}

For comparison with the mathematical model using quantum information theory to be explained below, we first use the optimal growth theory by Phelps and Pollak (1968) to obtain a utility function when consuming capital over time with exponential discounting.\\
First, we will outline optimal growth theory. In optimal growth theory, it is assumed that each generation lives, saves, and consumes in a single period. These periods are equally spaced and exist infinitely. All generations are assumed to be of the same size. The utility function $U$ of the current generation’s preferences is expressed as 

\begin{equation}
 U=u(C_{0} )+\alpha\delta u(C_{1} )+\alpha^{2}\delta u(C_{2} )+\dots, \quad  0<\delta<1,\quad 0<\alpha<1,
\end{equation}

where $C_{0}$ is the consumption of the current generation, $C_{1}$ is the consumption of the next generation, and so on. The period utility $u(C_{t})$ is the same function of current consumption, but the utility of consumption after $t$ periods is discounted by the factor $\delta\alpha^{t}$. The time factor $\delta$ is applied equally to all future generations regardless of timing and is an indicator of how much the current generation values the consumption of others relative to its own consumption. Imperfect altruism occurs when $0<\delta<1$, and perfect altruism occurs when $\delta=1$. All equations shown here are valid for any positive $\delta$, but for simplicity of explanation, we assume $\delta=1$ below.

\subsubsection{MATHEMATICAL REPRESENTATION USING OPTIMAL GROWTH THEORY}

Based on optimal growth theory, we set up utility functions for customers experiencing two subscription services and attempt to analyze them.

We consider two subscription services X and Y that are charged per unit time, and let $x_{t_{x}}$ and $y_{t_{y}}$ represent the remaining experienceable content quantities after consuming content up to service contract periods $t_{x}$ and $t_{y}$ respectively, with $\alpha$ as the discount rate and $u$ as the utility function. In particular, the utility function $u$ is given as

\begin{equation}
 u_x(x) = \log(x), \quad u_y(y) = \log(y).
\end{equation}

First, the problem of finding the utility function when a customer experiences one subscription service can be set up as follows as a dynamic programming problem, which is an infinite-period optimization problem:

\begin{equation}
\max_{\{x_{t_{x}}:t_{x}=0,1, \dots \}} \sum_{t_x=0}^\infty\alpha^{t_{x}}u_{x}(x_{t_x}-x_{t_x+1}) \notag \\
\end{equation}

\begin{align}
\text{Subject to:} \quad 
& x_{t_x} - x_{t_x + 1} \geq 0, \quad t_x = 0, 1, \dots \notag \\
& c_{x0} = x_0. \notag \\
\end{align}

From Hara and Kajii (2008), the solution to this is known to be

\begin{equation}
u_{x} (x_{t_{x}} - x_{t_{x}+1}) = \log(\alpha^{t_{x}} - \alpha^{t_{x}+1}).
\end{equation}

Next, the problem of finding the utility function when a customer experiences two subscription services can be set up as follows as a dynamic programming problem, which also is an infinite-period optimization problem:

\begin{equation}
\max_{\substack{\{x_{t_{x}}:t_{x}=0,1, \dots \} \\ \{y_{t_{y}}:t_{y}=0,1, \dots \}}}  \sum_{t_{x}=0}^\infty\alpha^{t_{x}}u_{x}(x_{t_x}-x_{t_x+1})+ \sum_{t_{y}=0}^\infty\alpha^{t_{y}}u_{y}(y_{t_y}-y_{t_y+1}) \notag \\
\end{equation}

\begin{align}
\text{Subject to:} \quad 
& x_{t_x} - x_{t_x + 1} \geq 0, \quad t_x = 0, 1, \dots \notag \\
& y_{t_y} - y_{t_y + 1} \geq 0, \quad t_y = 0, 1, \dots \notag \\
& c_{x0} = x_0, \notag \\
& c_{y0} = y_0. \notag \\
\end{align}

Here, because X and Y are independent, when we consider finding the maximum value as a dynamic programming problem of infinite-period optimization for each of X and Y separately, $u_{x}$ and $u_{y}$ become

\begin{equation}
u_{x} (x_{t_{x}} - x_{t_{x}+1}) = \log(\alpha^{t_{x}} - \alpha^{t_{x}+1})c_{x_{0}}, \quad 
u_{y} (y_{t_{y}} - y_{t_{y}+1}) = \log(\alpha^{t_{y}} - \alpha^{t_{y}+1})c_{y_{0}}.
\end{equation}

Then, since we are considering a mathematical model of subscription services in standard economics, we take into account exponential discounting. Here, if $k$ is the exponential discount rate, then $\alpha =e^{-k}$. For simplicity, we set $k=1$ and $\alpha =e^{-1}$.

Then, the customer satisfaction $V(T_{x},T_{y})$ when experiencing the subscription services up to contract periods $T_{x}$ and $T_{y}$ becomes

\begin{align}
V(T_x, T_y) &= \sum_{t_x = 0}^{T_x} e^{-t_x} \log(e^{-t_x}-e^{-t_x-1})c_{x0}
+ \sum_{t_y = 0}^{T_y} e^{-t_y} \log(e^{-t_y}-e^{-t_y-1})c_{y0} \notag \\
&= \frac{1}{(1 - e^{-1})^2} \Big[ 
(1 - e^{-1}) T_x e^{-T_x - 1} 
- e^{-1} (1 - e^{-T_x}) \notag \\
&\quad + (1 - e^{-1})(1 - e^{-T_x - 1}) \log(1 - e^{-1}) c_{x0} 
\Big] \notag \\
&\quad + \frac{1}{(1 - e^{-1})^2} \Big[ 
(1 - e^{-1}) T_y e^{-T_y - 1} 
- e^{-1} (1 - e^{-T_y}) \notag \\
&\quad + (1 - e^{-1})(1 - e^{-T_y - 1}) \log(1 - e^{-1}) c_{y0} 
\Big] \notag \\
&= V(T_x) + V(T_y).
\end{align}

Here, since $V(T_{x})$ and $V(T_{y})$ have the same functional form for X and Y, if we let $c_{x0}$ and $c_{y0}$ be ${C}$, then $V(t)$ becomes

\begin{equation}
 V(t)=\frac{1}{(1-e^{-1})^2}\Big[(1-e^{-1})te^{-t-1}-e^{-1}(1-e^{-t})+(1-e^{-1})(1-e^{-t-1})\log(1-e^{-1})C\Big].
\end{equation}

When displaying graphs of $V(t)$ by changing the magnitude of $C$, the result is as shown in Fig. 2.

The customer satisfaction $V(t)$ takes the form of a convex function, and as the consumed content quantity increases, the customer satisfactions $V(T_{x})$ and $V(T_{y})$ increase, and the maximum value shifts to the right in the direction of elapsed service time. Furthermore, for subscription services with larger total content quantities, the decrease in customer satisfaction magnitude tends to become slower as experience time increases. This means that as content quantity increases, customer satisfaction increases and the maximum value shifts rightward, so the customer’s experience time increases, and after customer satisfaction reaches its maximum, the customer gradually becomes bored with the subscription service and customer satisfaction decreases. It can be seen that the customer satisfaction $V(T_{x},T_{y})$ represents an indifference curve where customer satisfaction increases and then decreases as contract time increases.

\begin{figure}[H]
    \centering
    \includegraphics[width=0.6\linewidth]{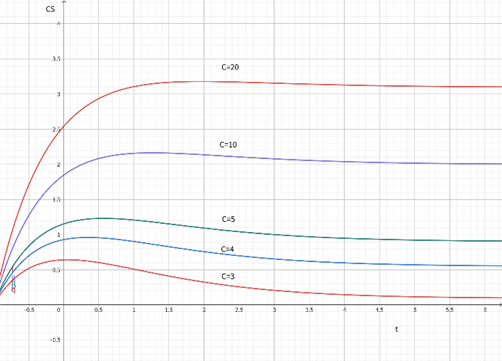}
    \caption{Graph of customer satisfaction V(t).}
    \label{fig:placeholder}
\end{figure}

The budget constraint equation for two subscription services X and Y is set up as

\begin{equation}
\ P=P_{x}T_{x}+P_{y}T_{y},
\end{equation}

where $P$, $P_{x}$, and $P_{y}$ are constant, with $P$ representing the budget, $P_{x}$ and $P_{y}$ representing fees per unit time, and $T_{x}$ and $T_{y}$ representing the contract times for the subscription services.

The condition for the time-discounted customer satisfaction value function to take an extreme value under the budget constraint equation is obtained by the Lagrange multiplier method. Letting $\lambda$ be the Lagrange multiplier, the Lagrangian $\Lambda$ is

\begin{equation}
\ \Lambda=V(T_{x},T_{y} )+\lambda(P-P_{x}T_{x}-P_{y}T_{y}). 
\end{equation}

From Fig. 4, The second partial derivatives of $\Lambda$ and the Hessian $H$ are

\begin{equation}
\frac{{\partial^2}\Lambda}{\partial T_{x}^2}=\frac{\partial^{2}V}{\partial T_{x}^2}>0,  \quad  
\frac{{\partial^2}\Lambda}{\partial T_{y}^2}=\frac{\partial^{2}V}{\partial T_{y}^2}>0,  \quad
\end{equation}

\begin{equation}
H =
\begin{vmatrix}
\frac{\partial^2 \Lambda}{\partial T_x^2} & \frac{\partial^2 \Lambda}{\partial T_x \partial T_y} \\
\frac{\partial^2 \Lambda}{\partial T_y \partial T_x} & \frac{\partial^2 \Lambda}{\partial T_y^2}
\end{vmatrix}
= \frac{\partial^2 \Lambda}{\partial T_x^2} \cdot \frac{\partial^2 \Lambda}{\partial T_y^2} > 0
\end{equation}

In Fig. 3, the intersection of the horizontal line $\lambda P_{x}$ and $\frac{\partial V}{\partial t}$ gives $\frac{\partial \Lambda}{\partial t}=0$, and since $\frac{\partial \Lambda}{\partial t}$ changes from negative to positive, it can be seen that the customer satisfaction $V(t)$ under budget constraints has a minimum value. That is, for customer satisfaction under budget constraints, as the total content quantity increases, the minimum value shifts to the right with respect to contract time each time the price is lowered. This represents that the lower the price, the longer it takes from when satisfaction reaches maximum until boredom begins.

\begin{figure}[H]
    \centering
    \includegraphics[width=0.6\linewidth]{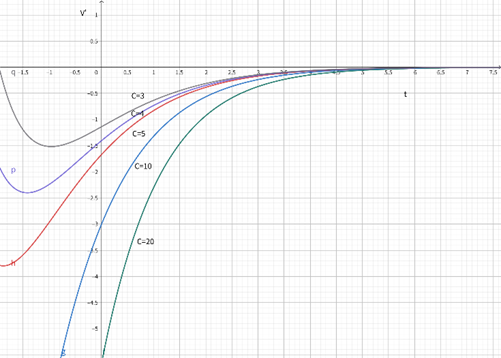}
    \caption{Graph of the first derivative of customer satisfaction V(t).}
    \label{fig:placeholder}
\end{figure}

\begin{figure}[H]
    \centering
    \includegraphics[width=0.6\linewidth]{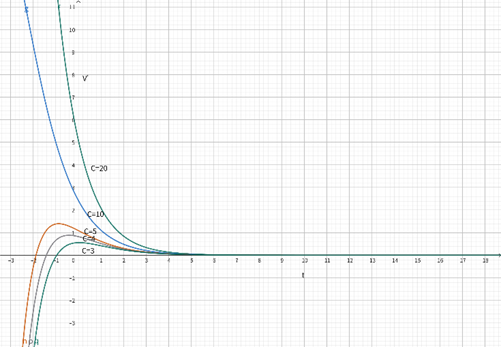}
    \caption{Graph of the second derivative of customer satisfaction V(t).}
    \label{fig:placeholder}
\end{figure}

\subsection{MATHEMATICAL MODEL OF INTANGIBLE GOODS}
\subsubsection{REPRESENTATION OF INTANGIBLE GOODS IN HILBERT SPACE}

According to Fukuda (2022), the value function of customer satisfaction for intangible goods that have socially diffused to a degree $x$ is

\begin{equation}
S_{merit}=p(\uparrow)\log\left(\frac{x}{p(\uparrow)}+1 \right), \quad S_{demerit}=p(\downarrow)\log\left(\frac{x}{p(\downarrow)}+1\right),
\end{equation}

\begin{equation}
p(\uparrow)=\frac{1}{N}\sum_{n=0}^{N-1}|\alpha_{n}|^2-\frac{2}{N^2-N}\sum_{m>n}|\alpha_{n}^*\alpha_{m}|\cos((\theta_m-\phi_m)-(\theta_n-\phi_n)),
\end{equation}

\begin{equation}
p(\downarrow)=1-p(\uparrow),
\end{equation}

\begin{equation}
\sum_{n=0}^{N-1}|\alpha_n|^2=t, \quad  (0\leq t\leq N)
\end{equation}

where $\sum_{m>n}$ denotes the sum of terms satisfying the condition $m>n$. Also, $p(\uparrow)$ represents the weight of emotional satisfaction that changes the repeat rate for intangible goods consisting of $N$ processes/functions, $\alpha_n $ is the magnitude of the impact felt by the customer for the $n$-th process/function, and the phase represents quality. $p(\downarrow)$ is the magnitude of the repeat rate. 

Also, assuming two intangible goods, let $\pi_1$ and $\pi_2$ be intangible goods of the same object that are perceived as satisfying. Considering the intangible goods $\pi_1+\pi_3$ that are formally perceived as satisfying by setting a mixture of two intangible goods, the weight of emotional satisfaction that changes the repeat rate of $\pi_1+\pi_3$ as a satisfaction factor, $p(\uparrow_1+\uparrow_3)$, is formally expressed as

\begin{equation}
p(\uparrow_1+\uparrow_3)=\mathrm{Tr}\rho P(\uparrow)=p(\uparrow_1)+p(\uparrow_3)+q(\uparrow_1+\uparrow_3).
\end{equation}

Replacing the second term on the right side,

\begin{equation}
p(\uparrow_1) = \sum_{n=0}^{l-1} |b_{n} \alpha_{n}|^2 
- \frac{2}{l^2 - l} \sum_{l-1 \geq m > n} |\alpha_{n}^{*} \alpha_m| 
\cos\left((\theta_m - \phi_m) - (\theta_n - \phi_n)\right)
\end{equation}

\begin{equation}
p(\uparrow_3) = \sum_{n=l}^{2N-1} |b_{n} \alpha_{n}|^2 
- \frac{2}{(2N - l)^2 - 2N + l} \sum_{2N-1 \geq m > n \geq l} 
|\alpha_{n}^{*} \alpha_m| 
\cos\left((\theta_m - \phi_m) - (\theta_n - \phi_n)\right)
\end{equation}

\begin{equation}
q(\uparrow_1 + \uparrow_3) = 
- \frac{2}{((2N - l) l)^2 - (2N - l) l} 
\sum_{2N - 1 \geq m \geq l > n} 
|\alpha_{n}^{*} \alpha_m| 
\cos\left((\theta_m - \phi_m) - (\theta_n - \phi_n)\right)
\end{equation}

where $\sum_{l-1\geq m>n}$ ,$\sum_{2N-1\geq m > n\geq l}$, and $\sum_{2N-1 \geq m \geq l>n}$ denote sums of terms where $m$ and $n$ satisfy the conditions $l-1\geq m>n$, $2N-1\geq m > n\geq l$, and $2N-1\geq m \geq l>n$ respectively, $P(\uparrow)$ is a positive operator-valued measure (POVM), and $|b_n|$ is the magnitude of the impact felt by the customer for the $n$-th process/function before experiencing the intangible goods. 

\subsubsection{MATHEMATICAL REPRESENTATION USING QUANTUM INFORMATION THEORY}

Considering subscription services charged per unit time, we obtain the consumer surplus by finding the condition under which the customer satisfaction value function takes an extreme value under budget constraints. Setting two subscription services as X and Y, the budget constraint equation is set up as

\begin{equation}
P=P_{x}T_{x}+P_{y}T_{y}, 
\end{equation}

where $P$, $P_x$, and $P_y$ are constant, with $P$ representing the budget, $P_x$ and $P_y$ representing fees per unit time, and $T_x$ and $T_y$ representing the contract times for the subscription services.

Here, based on Takahashi (2005), we consider the time discounting of the customer satisfaction value function when there is correlation in the customer’s perception of two subscription services. According to Fechner’s law, the sensory magnitude is the logarithm of stimulus intensity, so it becomes

\begin{equation}
 \tau(D)=\alpha \log(1+\beta D),
\end{equation}
 
where $\alpha$ and $\beta$ are some constants, $\tau$ is subjective time, and $D$ is objective elapsed time (Fechner 1860).

The value function $S_i$ when two stimuli are felt is, from exponential discounting,

\begin{equation}
S_i=A\exp(-k_i(\tau_1+\tau_2)),
\end{equation}

\begin{equation}
\tau_{1}(D_{1})=\alpha_{1}\log(1+\beta_{1}D_{1})=\log(1+\beta_{1}D_{1})^{\alpha_{1}},
\end{equation}

\begin{equation}
\tau_{2}(D_{2})=\alpha_{2}\log(1+\beta_{2}D_{2})=\log(1+\beta_{2}D_{2})^{\alpha_{2}},
\end{equation}

which becomes

\begin{equation}
S_i=\frac{A}{(1+\beta_{1}D_{1})^{k_{i}\alpha_{1}}(1+\beta_{2}D_{2})^{k_{i}\alpha_{2}}},
\end{equation}

where $A$, $k_i$, $\alpha_1$, $\beta_1$, $\alpha_2$, and $\beta_2$ are some constants, $\tau_1$ and $\tau_2$ are subjective times, and $D_1$ and $D_2$ are objective elapsed times.

Also, the quantum mutual information $S(x:y)$ with two components $x$ and $y$ is defined as

\begin{equation}
S(x:y)=S(x)+S(y)-S(x,y), 
\end{equation}

where $S(x)$ and $S(y)$ are von Neumann entropies, and $S(x,y)$ represents the joint entropy of the two components $x$ and $y$, defined as

\begin{equation}
S(x)=-\mathrm{Tr}\rho_{x}\log\rho_{x}=-\mathrm{Tr}\sum_{i}p_{i}\log p_{i}|i><i|=-\sum_{i}p_{i}\log p_{i},
\end{equation}

\begin{equation}
S(y)=-\mathrm{Tr}\rho_{y}\log\rho_y=-\mathrm{Tr}\sum_i q_{i}\log q_i  |i><i|=-\sum_{i} q_i  \log q_i,
\end{equation}

\begin{equation}
S(x,y)=-\mathrm{Tr}\rho_{xy} \log\rho_{xy}=-\mathrm{Tr}\sum_{i,j}p_{i}q_{j}\log p _{i} q _{j}  |ij><ij|=-\sum_{i,j} p _{i}  q_{j}  \log{p}_{i}  q_{j},
\end{equation}

\begin{equation}
\rho_{x}=\sum_i p_{i} |i><i|, \quad  
\rho_y=\sum_i q_{i}  |i><i|,\quad  
\rho_{xy}=\rho_{x}\otimes\rho_y=\sum_{i,j} p_i  q_j  |ij><ij|.
\end{equation}

With these in mind, and based on Eq. (17), the customer satisfaction value function $S(\uparrow_1+\uparrow_3)$ discounted from the customer’s present value when the customer experiences two subscription services for contract times $T_{x}$ and $T_{y}$ is defined as

\begin{equation}
S(\uparrow_{1}+\uparrow_{3})=\frac{S(\uparrow_{1})}{(1+\beta_{1}T_x)^{k_{1}\alpha_1}}+\frac{S(\uparrow_3)}{(1+\beta_{2}T_y)^{k_{2}\alpha_{2}}} 
-\frac{S(\uparrow_1,\uparrow_3)}{(1+\beta_{1}T_x)^{k\alpha_1} (1+\beta_2T_y)^{k\alpha_2}},
\end{equation}

where

\begin{equation}
S(\uparrow_1)=p(\uparrow_1) \log\left( \frac{x}{p(\uparrow_1)}+1 \right), \quad  
S(\uparrow_3)=p(\uparrow_3) \log\left( \frac{x}{p(\uparrow_3)}+1 \right), \quad 
S(\uparrow_1,\uparrow_3)=|q(\uparrow_1+\uparrow_3)| \log\left( \frac{x}{|q(\uparrow_{1}+\uparrow_{3})|}+1\right). 
\end{equation}

Setting $k=k_1=k_2=\alpha_1=\beta_1=\alpha_2=\beta_2=1$ for simplicity, the partial derivatives of $S(\uparrow_1+\uparrow_3)$ become

\begin{equation}
 \frac{\partial S(\uparrow_1+\uparrow_3)}{\partial T_x}=\frac{1}{(1+T_x)^2}\left(\frac{S(\uparrow_1,\uparrow_3)}{1+T_y}-S(\uparrow_1)\right), \quad  
  \frac{\partial S(\uparrow_1+\uparrow_3)}{\partial T_y}=\frac{1}{(1+T_y)^2}\left(\frac{S(\uparrow_1,\uparrow_3)}{1+T_y}-S(\uparrow_3)\right). 
\end{equation}

When $S(\uparrow_1,\uparrow_3)=0$,

\begin{equation}
\frac{\partial S(\uparrow_1+\uparrow_3)}{\partial T_x}<0, \quad  \frac{\partial S(\uparrow_1+\uparrow_3)}{\partial T_y}<0, 
\end{equation}

always holds. Thus, we can see that $S(\uparrow_1+\uparrow_3)$ represents an indifference curve where customer satisfaction decreases as contract time increases. Also, when $S(\uparrow_1,\uparrow_3) \neq 0$ and setting $p(\uparrow_1)=p(\uparrow_3)=0.15$ and $q(\uparrow_1+\uparrow_3)=0.35$ as average weights, the partial derivative of $S(\uparrow_1+\uparrow_3)$ changes from positive to negative as contract time increases, so we can see that when $S(\uparrow_1+\uparrow_3)$ exceeds a certain contract time, it becomes a convex indifference curve that changes from increasing to decreasing.

Next, referring to Tanaka (2013) and Hara (2018), the condition for the time-discounted customer satisfaction value function to take an extreme value under the budget constraint equation is obtained by the Lagrange multiplier method. Letting $\lambda$ be the Lagrange multiplier, the Lagrangian $\Lambda$ is

\begin{equation}
 \Lambda = S(\uparrow_1+\uparrow_2)+\lambda(P-P_{x}T_{x}-P_{y}T_{y}). 
\end{equation}

The second partial derivatives of $\Lambda$ and the Hessian $H$ are

\begin{equation}
 \frac{\partial^2 \Lambda}{\partial T_x^2}=\frac{2}{(1+T_x)^3}    \left(S(\uparrow_1)-\frac{S(\uparrow_1,\uparrow_3)}{1+T_y}\right), \quad  
 \frac{\partial^2 \Lambda}{\partial T_y^2}=\frac{2}{(1+T_y)^3}    \left(S(\uparrow_3)-\frac{S(\uparrow_1,\uparrow_3)}{1+T_x}\right), 
\end{equation}

\begin{equation}
H =
\begin{vmatrix}
\frac{\partial^2 \Lambda}{\partial T_x^2} & \frac{\partial^2 \Lambda}{\partial T_x \partial T_y} \\
\frac{\partial^2 \Lambda}{\partial T_y \partial T_x} & \frac{\partial^2 \Lambda}{\partial T_y^2}
\end{vmatrix}
= \frac{4}{(1+T_x)^3(1+T_y)^3} \left(
S(\uparrow_1) S(\uparrow_3)
- \left(
\frac{S(\uparrow_1)}{1+T_x}
+ \frac{S(\uparrow_3)}{1+T_y}
\right)
S(\uparrow_1, \uparrow_3)
\right).
\end{equation}

Therefore, when $S(\uparrow_1,\uparrow_3)=0$, $\frac{\partial^2\Lambda}{\partial T_x^2}>0$, $\frac{\partial^2\Lambda}{\partial T_y^2}>0$, and $H>0$
always hold. Thus, when $\frac{\partial\Lambda}{\partial T_x}=\frac{\partial\Lambda}{\partial T_y}=0$, $\Lambda$ has a minimum value. Also, when $S(\uparrow_1,\uparrow_3)\neq0$, similarly setting average weights, the second partial derivatives of $\Lambda$ and the Hessian H change from negative to positive as contract time increases, and after a certain contract time, they take a minimum value.

\section{CONSUMER SURPLUS}
Here, considering only when $\Lambda$ takes a minimum value, we obtain the following by the Lagrange multiplier method. Let the time-discounted customer satisfaction value function under budget constraints be collectively denoted as $V(T_x,T_y)=S(\uparrow_1+\uparrow_3)=V$. Then,

\begin{equation}
 \frac{\partial\Lambda}{\partial T_x}=\frac{\partial V}{\partial T_x} -\lambda P_x=0, \quad  
 \frac{\partial\Lambda}{\partial T_y}=\frac{\partial V}{\partial T_y} -\lambda P_y=0. 
\end{equation}

Eliminating $\lambda$ from this result,

\begin{equation}
\frac{\frac{\partial V}{\partial T_x}}{\frac{\partial V}{\partial T_y}}=\frac{P_x}{P_y}.
\end{equation}

Therefore, the minimum condition of the time-discounted customer satisfaction value function $V$ under budget constraints is, similarly to the utility maximum condition under budget constraints in conventional microeconomics, that the marginal rate of substitution equals the relative price. From this conditional equation and the budget constraint equation, $T_x$ and $T_y$ can be expressed as functions of $P_x$ and $P_y$.

Next, we find the contract time for each subscription service when minimizing expenditure with the time-discounted customer satisfaction value function $V$ held constant. From the budget constraint equation, the expenditure $m$ is obtained as

\begin{equation}
 m=P_{x}T_{x}+P_{y}T_{y}.
\end{equation} 

Taking the partial derivative of this equation with respect to $P_x$,

\begin{equation}
 \frac{\partial m}{\partial P_x}=T_x+P_{x}\frac{\partial T_x}{\partial P_x}+P_{y}\frac{\partial T_y}{\partial P_x}.
\end{equation} 

Also, taking the partial derivative of the constant time-discounted customer satisfaction value function $V$ with respect to $P_x$ gives

\begin{equation}
 \frac{\partial V}{\partial T_x}\frac{\partial T_x}{\partial P_x}+\frac{\partial V}{\partial T_y}\frac{\partial T_y}{\partial P_x}=0.
\end{equation} 

Thus, the conditional equation for the contract time of the subscription service when expenditure is minimized and the time-discounted customer satisfaction value function is minimized is

\begin{equation}
 \frac{\partial m}{\partial P_x}=T_x, 
\end{equation} 

\begin{equation}
 P_{x}\frac{\partial T_x}{\partial P_x}+P_{y}\frac{\partial T_y}{\partial P_{x}}=0. 
\end{equation} 

We can similarly obtain

\begin{equation}
 \frac{\partial m}{\partial P_y}=T_y,
\end{equation}

\begin{equation}
 P_{x}\frac{\partial T_x}{\partial P_y}+P_{y}\frac{\partial T_y}{\partial P_{y}}=0.         
 \end{equation}
 
Thus, in the expenditure minimization condition of the second equations in Eqs. (45) and (47),

\begin{equation}
 \frac{\partial T_x}{\partial P_x}<0, \quad  \frac{\partial T_y}{\partial P_y}<0,
\end{equation}
 
and therefore,

\begin{equation}
 \frac{\partial T_y}{\partial P_x}>0, \quad  \frac{\partial T_x}{\partial P_y}>0,
\end{equation}

which shows that subscription services X and Y are mutual substitutes.

Thus, the change in consumer surplus $\Delta CS$ when changing the per-unit-time prices of the two subscription services X and Y from $P_x$ to $P_x'$ and from $P_y$ to $P_y'$ is expressed by the following defining equation, while satisfying (45) and (47) under the condition that X and Y are mutual substitutes:

\begin{equation}
\Delta CS = \int_{P_x}^{P_x'} T_x \, dP_x + \int_{P_y}^{P_y'} T_y \, dP_y.
\end{equation}

If the income effect can be ignored, $\Delta CS$ becomes an indicator of economic welfare because it does not depend on the order in which the prices of the two subscription services change.

Also, considering the case of the mathematical model using quantum information theory, after the time-discounted customer satisfaction value function is minimized,

\begin{equation}
\frac{\frac{\partial V}{\partial T_x}}{\frac{\partial V}{\partial T_y}} \neq \frac{P_x}{P_y},
\end{equation}

must be satisfied while adjusting the coefficients of the basis that constitute the weight of emotional satisfaction that changes the time-discounted customer satisfaction repeat rate, so as to maximize $\Delta CS$.

\section{DESIGN OF CUSTOMIZED CUSTOMER EXPERIENCE}
We describe the design of customized customer experiences in subscription services. Because the mathematical model using standard economics cannot express customization, we consider the mathematical model using quantum information theory. First, we obtain the coefficients of the basis from the minimum condition as a function of time for the perception $P(\uparrow)$ of the supply and demand process of the subscription service when the value function becomes a minimum at a certain contract time, and reduce the contract time at which this minimum value occurs. At this time, it is possible to express the conditional equations for customized nudges and the timing of intervention and repeat rate that cannot be expressed in standard economics. After this, as the customer satisfaction value function increases under budget constraints after the contract time at which this minimum value occurs, obtaining the coefficients of the basis as a function of time for the perception $P(\uparrow)$ of the supply and demand process of the subscription service, within the range satisfying Eq. (51), for the two subscription services being supplied and demanded, constitutes the design of customized customer experience for individual customers.

\section{Results}
In the customer satisfaction value function of the mathematical model of subscription services in this research, in the case of the mathematical model using optimization problems in standard economics, it becomes a convex function with respect to time, and as the total content quantity increases, the customer satisfaction value function increases and the maximum value shifts in the time direction. In the case of the mathematical model using quantum information theory, when there is no interaction between the two subscription services, the customer satisfaction value function becomes a decreasing function with respect to time, and when there is interaction, it becomes a convex function with respect to time. The content quantity in the mathematical model using optimization problems in standard economics corresponds to the weight of the emotional satisfaction term in the mathematical model using quantum information theory. When there are budget constraints, it was found that both the standard economics and quantum information theory mathematical models have minimum values. In particular, in the mathematical model using quantum information theory, it was possible to express the conditional equations for customized nudges and the timing of intervention and repeat rate that cannot be expressed in standard economics. Therefore, the mathematical model using quantum information theory has a wider range of expression than the mathematical model using optimization problems in standard economics. From this, it was suggested that the customer satisfaction value function derived based on quantum information theory may represent the general outline of customer satisfaction for customers experiencing subscription services.

As a property of consumer surplus under budget constraints, the prospect emerged that when the time-discounted customer satisfaction value function is minimized under temporal budget constraints, it becomes a mathematical model that maximizes consumer surplus.

This suggested the manifestation of new methods from a behavioral economics perspective that do not exist in conventional approaches in marketing, social policy, or other areas.

Table 2 shows a comparison summarizing the mathematical results and application fields between the existing expected utility theory model in standard economics and the model in this research.

\begin{table}[htbp]
\centering
\caption{Comparison between Standard Economics and Quantum Information Theory Models}
\begin{tabular}{|p{4.8cm}|p{4.8cm}|p{4.8cm}|}
  \hline
  & Standard Economics Model & Quantum Information Theory Model \\
  \hline
  Properties of consumer surplus under budget constraints & 
  When utility is maximized under budget constraints on commodity quantity, consumer surplus is maximized & 
  When the time-discounted customer satisfaction value function is minimized under temporal budget constraints, consumer surplus is maximized \\
  \hline
  Application fields & 
  Marketing, social policy, etc. & 
  Manifestation of new methods from a quantum information perspective that do not exist in the conventional methods mentioned on the left \\
  \hline
\end{tabular}
\end{table}

\section{Discussion}
The following is a discussion of subscription services. As a case where budget constraints exist, when the time-discounted customer satisfaction value function in a customer’s experience of two subscription services is mathematically expressed as Eq. (13), the fact that this time-discounted customer satisfaction value function is minimized means that the weight of the emotional satisfaction term that changes the repeat rate in Eq. (13) is minimized as a function of time. In particular, the gain of the customer satisfaction value function decreases in the direction of becoming minimum, and changes in the direction of increasing loss. From this, the customer repeats to avoid loss and obtain gain. Then, due to status quo bias, the customer tries to increase gain information by collecting only reasons that justify repeating, does not collect loss information, and if nothing particular happens, remains convinced by and satisfied with the intangible goods supply and demand experience and continues to repeat.

As a result, as stated in Janzer (2017), when customers first pay fees for subscription services, customer satisfaction increases and they continue to repeat. Therefore, after a certain contract time when the customer satisfaction value function is minimized for subscription services under budget constraints, as the customer satisfaction value function increases by improving quality and other measures, it is simultaneously necessary to adjust Eq. (51) to design prices so that consumer surplus becomes at least $\Delta CS$, and to increase $\Delta CS$. Also, if the contract time until taking this minimum value is reduced and payment is made earlier, it is likely that the value function will shift to increase at an earlier contract time. 

Furthermore, the timing when this minimum value is reached is when the behavior change rate is high, so it is likely that if interventions such as nudges are made with this timing, the repeat rate will increase. In this respect, since consumer surplus described by quantum information theory can be made larger than consumer surplus described by standard economics, it is likely that the mathematical model of subscription services described by quantum information theory can achieve greater economic welfare. Also, there is a possibility that infinite-period optimization problems solved by dynamic programming can be formulated by replacing them with quantum information theory, and that optimization problem algorithms can be implemented on quantum computers.

\section{Summary}
In this research, we set up and analyzed a mathematical model of subscription services from a time discounting perspective based on quantum information theory. By finding correspondences between the minimum condition under budget constraints of the defined value function and the laws of behavioral economics, the possibility increased that this could be a candidate for a mathematical design model of subscription services and a mathematical design model of customized customer experience. Furthermore, the possibility increased that the mathematical model of subscription services described by quantum information theory can achieve greater economic welfare than description by standard economics. Also, the possibility increased that optimization problem algorithms can be implemented on quantum computers.

\section*{Appendix}
The following outlines POVM, as per Hayashi (2004). Measurement is necessary to obtain information about nanoscale objects in quantum systems. The conditions of such a system, such as the direction of electron spin, is called its “state.” This state is called a density matrix, represented by a positive semidefinite Hermitian matrix $\rho$, and satisfies the properties

\begin{equation}
\mathrm{Tr}\rho=1, \quad  <n|\rho|n> \geq 0. 
\end{equation}

Also, the positive semidefinite Hermitian matrix $P(\omega)$ representing the measurement to obtain information of the system’s data $\omega$ is called a POVM and satisfies the properties

\begin{equation}
\sum_{\omega}P(\omega)=I, \quad  <n|P(\omega)|n> \geq 0, 
\end{equation}

where $<n-1|$ is a unit vector with $1$ in the first row and $n$-th column, $|n-1>$ is a unit vector with 1 in the $n$-th row and first column, and $I$ is the identity matrix.

The probability $p(\omega)$ that datum $\omega$ is obtained is given as

\begin{equation}
p(\omega)=\mathrm{Tr}\rho P(\omega).
\end{equation}

This is

\begin{equation}
\sum_{\omega}p(\omega)=\sum_{\omega}\mathrm{Tr}\rho P(\omega)=\mathrm{Tr}\rho\sum_{\omega}P(\omega)=\mathrm{Tr}\rho I=\mathrm{Tr}\rho=1,
\end{equation}

\begin{equation}
\mathrm{Tr}\rho P(\omega)=\sum_{n}<n|\rho P(\omega)|n>=\sum_{n}<n|\rho|n><n|P(\omega)|n> \geq 0,
\end{equation}

and we can see that $p(\omega)$ is a probability.

\section*{Acknowledgments}
The author thanks Prof.T.Kurata of RIHSA at AIST, Prof.N.Nishino, and Assoc.prof.T.Hara of the University of Tokyo for insightful discussions.

\bibliographystyle{unsrt}  


\begin{thebibliography}{1}


\bibitem{Ascarza}
Ascarza, E., Neslin, S. A., Netzer, O., Anderson, Z., Fader, P. S.,Gupta, S., Bruce G. S. Hardie, A. Lemmens, B. Libai, David Neal, F. Provost,  \& Schrift, R. (2018). In pursuit of enhanced customer retention management: Review, key issues, and future directions. Customer Needs and Solutions, 5(1), 65–81. \url{https://doi.org/10.1007/s40547-017-0080-0}

\bibitem{Cheon}
Cheon, T. \& Takahashi, T. (2010). Interference and inequality in quantum decision theory. Physics Letters A, 375(2), 100‒104. \url{https://doi.org/10.1016/j.physleta.2010.10.063}

\bibitem{Dazai}
Dazai, U. (2022). The characteristics of customer satisfaction and subscription service usage. Quarterly Journal of Marketing, 41(3), 18–29.  \url{https://doi.org/10.7222/marketing.2022.004}

\bibitem{Fechner}
Fechner. G. (1860). Elemente der Psychophysik, Leipzing, Germany: Breitkopf \& Härtel

\bibitem{Fukuda 2022}
Fukuda, M. (2022). Proposal for reformulation of behavioral economic concepts centered on intangible goods by quantum information theory: Feasibility of mathematical design of customer experience. Journal of Behavioral Economics and Finance, 15, 10–21. \url{https://doi.org/10.11167/jbef.15.10}

\bibitem{Fukuda 2023}
Fukuda, M. Mathematical representation of bias and nudge centered on intangible goods by quantum information theory, Poster presented at The 10th InternationalCongress on Industrial and Applied Mathematics (ICIAM 2023), 20-25 Aug. 2023, Tokyo.

\bibitem{Fukuda 2024}
Fukuda, M. (2024). Mathematical representation of bias and nudges centered on intangible goods using quantum information theory. \url{https://arxiv.org/abs/2411.08046}

\bibitem{Hara 2008}
Hara, C., \& Kajii, A. (2008). Mathematics for economics. Institute of Economic Research, Kyoto University. \url{https://www.hara.kier.kyoto-u.ac.jp/NoteByInami3.pdf}.

\bibitem{Hara 2018}
Hara, C. (2018). Advanced microeconomics lecture notes. Institute of Economic Research, Kyoto University. 
\url{http://www.hara.kier.kyoto-u.ac.jp/AdvMicro18LectureNotes.pdf}.

\bibitem{Hayashi}
Hayashi, M. (2004). Introduction to quantum information theory. Saiensusha, Tokyo.

\bibitem{Janzer}
Janzer, A. H. (2017). Subscription Marketing: Strategies for Nurturing Customers in a World of Churn (Second Edition). Cuesta Park Consulting.

\bibitem{Kanuri}
Kanuri, V. K., \& Andrews, M. (2019). The unintended consequence of price-based service recovery incentives. Journal of Marketing, 83(5), 57–77. \url{https://www.doi.org/10.1177/0022242919859325}

\bibitem{McCarthy}
McCarthy, D. M., Fader, P. S., \& Hardie, B. G. (2017). Valuing  subscription-based businesses using publicly disclosed customer data. Journal of Marketing, 81(1), 17 – 35. \url{https://www.doi.org/10.1509/jm.15.0519}

\bibitem{Micken}
Micken, K. S., Roberts, S. D., \& Oliver, J. D. (2020). The digital continuum: The influence of ownership, access, control, and cocreation on digital offerings. AMS Review, 10, 98–115. \url{https://www.doi.org/10.1007/s13162-019-00149-5}

\bibitem{Nielsen}
Nielsen, M. A., \& Chuang, I. L. (2000). Quantum Computation and Quantum Information. Cambridge University Press.

\bibitem{Okuno}
Okuno, M. (2011). Microeconomics. University of Tokyo Press.

\bibitem{Phelps}
Phelps, E. S., \& Pollak, R. A. (1968). On second-best national saving and game equilibrium growth. Review of Economic Studies, 35, 185–199. \url{https://doi.org/10.2307/2296547}

\bibitem{Tanaka}
Tanaka, Y. (2021). Microeconomics. Faculty of Economics, Doshisha University. 
 \url{https://opencourse.doshisha.ac.jp/opc/attach/page/OPENCOURSE-PAGE-JA-316/152074/file/shokyumicrotogo.pdf}

\bibitem{Thaler}
Thaler, R. H., \& Sustein, C. R. (2008). Nudge: Improving Decisions About Health, Wealth and Happiness. Yale University Press.

\bibitem{Takahashi 2013}
Takahashi, T. (2013). Rationality in quantum decision theory. Kagaku Tetsugaku, 46(2), 17–30. \url{https://doi.org/10.4216/jpssj.46.17}

\bibitem{Takahashi 2005}
Takahashi, T. (2005). Loss of self-control in intertemporal choice may be attributable to logarithmic time-perception. Medical Hypothesis, 65(4), 691‒693. \url{https://doi.org/10.1016/j.mehy.2005.04.040}

\bibitem{Tversky}
Tversky, A., \& Kahneman, D. (1992). Advances in prospect theory: Cumulative representation of uncertainty. Journal of Risk and Uncertainty, 5, 297‒323. \url{https://doi.org/10.1007/BF00122574}

\bibitem{Yukalov}
Yukalov, V. I., \& Sornette, D. (2017). Quantum probabilities as behavioral probabilities. Entropy, 19(3), 112. \url{https://doi.org/10.3390/e19030112}


\end{thebibliography}

\end{document}